\documentclass[twocolumn,showpacs,preprintnumbers,amsmath,amssymb,superscriptaddress]{revtex4}

\usepackage{graphicx}
\usepackage{dcolumn}
\usepackage{bm}

\begin{document}


\title{Coherent Manipulation and Decoherence of S=10 Single-Molecule Magnets}
\author{Susumu Takahashi}
\email{susumu@iqcd.ucsb.edu}
\affiliation{Department of Physics and Center for Terahertz Science and Technology, University of California, Santa Barbara, California 93106}%

\author{Johan van Tol}
\affiliation{National High Magnetic Field Laboratory, Florida State University, Tallahassee Florida  32310}%

\author{Christopher C. Beedle}%
\affiliation{Department of Chemistry and Biochemistry, University of California, San Diego, La Jolla, California 92093}%

\author{David N. Hendrickson}%
\affiliation{Department of Chemistry and Biochemistry, University of California, San Diego, La Jolla, California 92093}%

\author{Louis-Claude Brunel}
\thanks{Present address: Center for Terahertz Science and Technology, University of California, Santa Barbara, CA}
\affiliation{National High Magnetic Field Laboratory, Florida State University, Tallahassee Florida  32310}%

\author{Mark S. Sherwin}%
\affiliation{Department of Physics and Center for Terahertz Science and Technology, University of California, Santa Barbara, California 93106}%

\date{\today}

\begin{abstract}
We report coherent manipulation of S=10 Fe$_{8}$ single-molecule
magnets. The temperature dependence of the spin decoherence time
$T_2$ measured by high frequency pulsed electron paramagnetic
resonance indicates that strong spin decoherence is dominated by
Fe$_{8}$ spin bath fluctuations. By polarizing the spin bath in
Fe$_{8}$ single-molecule magnets at magnetic field $B$ = 4.6 T and
temperature $T$ = 1.3 K, spin decoherence is significantly
suppressed and extends the spin decoherence time $T_2$ to as long
as 712 ns. A second decoherence source is likely due to
fluctuations of the nuclear spin bath. This hints that the spin
decoherence time can be further extended via isotopic substitution
to smaller magnetic moments.
\end{abstract}

\maketitle

%
%

Single-molecule magnets (SMMs) behave like nanoscale classical
magnets at high temperatures~\cite{Gatteschi}. The quantum
mechanical nature of SMMs emerges at low temperatures with
behaviors like quantum tunnelling of magnetization
(QTM)~\cite{friedman96, thomas96, sangregorio97}, quantum phase
interference of two tunnelling paths~\cite{wernsdorfer99,
ramsey08} and the observation of discrete transitions in electron
paramagnetic resonance (EPR) and optical
spectroscopy~\cite{Gatteschi, caneschi91, barra96, hill03,
mukhin01}. As quantum magnets based on solid state systems, SMMs
form a unique class of materials that have a high-spin, and their
spin state and interaction can be easily tuned by changing
peripheral organic ligands and solvate molecules. Because the
molecules within the crystal lattice of SMMs interact very weakly
with each other, properties of a single SMM can be deduced from
measurements of a macroscopic ensemble.

Although quantum phenomena observed in SMMs have been investigated
extensively, couplings between SMMs and their environment are
still poorly understood. Coupling to the environment results in
decoherence, which must be understood to optimize SMMs for
proposed applications to dense and efficient quantum memory,
computing, and molecular spintronics devices~\cite{Leuenberger01,
bogani08}. With pulsed EPR, it is possible to directly measure
spin relaxation times. However measurement of pulsed EPR for
high-spin SMMs has been extremely challenging due to strong spin
decoherence~\cite{deloubens08}. To our knowledge, no direct
measurements of the spin relaxation times have been reported for
single crystals of SMMs. Various investigations of spin dynamics
have led to the {\it estimate} that the lower bound of spin
decoherence time ($T_2$) in high-spin SMMs is on the order of
nanoseconds~\cite{hill03, deloubens08, delbarco04}. Decoherence of
SMMs must therefore be suppressed in order to measure $T_2$. One
approach to reduce spin decoherence has been to reduce the number
of spins in the bath. Molecule-based magnets have been studied
employing highly dilute solutions. Thereby, minimizing
intermolecular interactions leads to an increase of
$T_2$~\cite{ardavan07, schlegel08}. However, these compounds may
have different quantum and magnetic properties from those of
single crystals. Another approach has been to reduce the
fluctuations within spin baths. When the spin bath is fully
polarized, the spin bath fluctuations are completely eliminated.
For dilute, isotropic spins on nitrogen-vacancy (NV) centers in
diamond, spin decoherence due to electron spins has been quenched
by complete polarization (\textgreater~99\%) of the electron spin
baths with application of high magnetic fields and low
temperatures~\cite{takahashi08}.

\begin{figure*}
\includegraphics[width=175 mm] {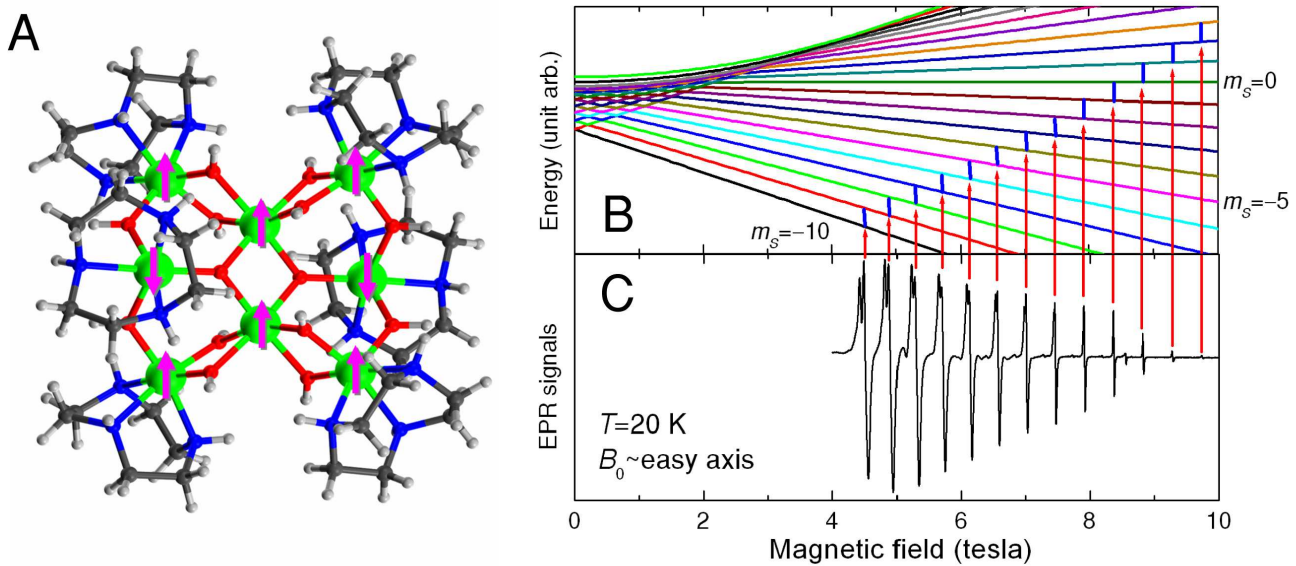}
\caption{\label{fig:cwEPR} (A) Schematic diagram of Fe$_8$
molecule (Fe: Green, O: Red, N: Blue, C: Gray and H: White). The
Fe$_8$ molecule consists of eight Fe(III) ($S=5/2$) ions which
couple to each other to form an $S=10$ ground state. (B) Energy
diagram as a function of the magnetic field, $\bm B_0$ while the
magnetic field is applied along the easy axis. The diagram is
calculated using Equation (1). (C) cw EPR spectrum at 20 K taken
by sweeping magnetic fields from 4 T to 10 T. The magnetic field
is applied along the easy-axis within 10 degrees. Corresponding
EPR transitions are indicated by blue solid lines in Fig. 1(B).
The spectrum also shows fine structures which are more pronounced
in the transitions of high $|m_S|$ values. These were also seen in
previous studies~\cite{Gatteschi}.  A resonance at 8.5 T is a
spurious signal from impurities in the sample holder.}
\end{figure*}

This paper presents measurements of the spin decoherence time
$T_2$ of single crystals of the $S=10$ SMM
{[Fe$_8$(O)$_2$(OH)$_{12}$(C$_6$H$_{15}$N$_3$)$_6$]Br$_7$(H$_2$O)}Br$\cdot$8H$_2$O,
abbreviated Fe$_8$~\cite{Wieghardt84}. pulsed EPR spectroscopy at
240 GHz and 4.6 T is used to measure $T_2$. The energy difference
between $m_S=-10$ and $m_S=-9$ states is 11.5 K, so Fe$_8$ SMMs
are almost completely polarized to the $m_S=-10$ lowest lying
state below 1.3 K. At 1.3 K, $T_2$ was measured to be 712 ns. Upon
raising the temperature to 2 K, the $T_2$ decreased by nearly 1
order of magnitude. As temperature increases, so do fluctuations
of the SMM spin bath. In order to describe the temperature
dependence, we extend a model of spin decoherence time for a
$S=1/2$ system to a system with arbitrary $S$. Good agreement
between the data and the model for Fe$_8$ SMMs strongly supports
the spin decoherence caused by fluctuations of Fe$_8$ SMMs spin
bath. We also show that there exist other decoherence sources
likely caused by proton and $^{57}$Fe nuclear magnetic moments
which limit the spin decoherence time $T_2$ $\sim$ 1 $\mu$s.

%
%
The magnetic properties of the $S=10$ Fe$_8$ SMM are
well-described by the following spin Hamiltonian,
\begin{eqnarray}
H = \mu_{B}g{\bm S}\cdot{\bm B_0}+ DS_z^2 + E(S_x^2-S_y^2),
\end{eqnarray}
where $\mu_B$ is the Bohr magneton, $\bm B_0$ is the magnetic
field and $S$ are the spin operators. The parameter $g$ = 2.00 is
an isotropic g-factor and $D$ = $-6.15$ GHz and $E$ = 1.14 GHz
represent the second order anisotropy constants~\cite{Gatteschi}.
The large spin ($S=10$) and the negative $D$ value lead to a large
energy barrier between the spin-up and spin-down states. Higher
order terms are not included here. As shown in the energy diagram
as a function of the magnetic field given in Fig.~1(B), at low
magnetic fields there are many energy level anti-crossings which
are the origin of quantum tunneling of magnetization
~\cite{Gatteschi, friedman96, thomas96}. On the other hand, in a
high magnetic field regime, where the magnetic field is aligned
along the easy axis and is higher than 4.3 T, there are no level
anti-crossings (see Fig.~1(B)).

%
%
Experiments were performed with the 240 GHz cw and pulsed EPR
spectrometer at the National High Magnetic Field Laboratory
(NHMFL), Tallahassee FL, USA. For this study, we employed a
superheterodyne quasioptical bridge with a 40 mW solid-state
source. In order to enable {\it in situ} rotation of the sample
relative to the applied magnetic field, we employed a rotating
sample holder mounted with its axis perpendicular to a 12.5 T
superconducting solenoid. A detailed description of the setup is
given elsewhere~\cite{morley08, vantol05}. Measurements were
performed on Fe$_8$ single crystal samples whose magnetic
anisotropic axis, called the easy axis, was identified by X-ray
diffraction measurements. Fig.~1(C) shows the cw EPR spectrum at
20 K while the magnetic field was applied along the easy axis
within 10 degrees. The applied microwave and field modulation
intensities were carefully tuned so as not to distort the EPR
lineshape. As indicated in Fig.~1(B) and (C), the spectrum shows
EPR transitions ranging from $m_S$ = $-10$ $\leftrightarrow$ $-9$
at 4.6 T to $m_S$ = 2$\leftrightarrow$3 at 9.7 T. In addition,
much weaker EPR signals from the $S=9$ excited state transitions
were observed ~\cite{zipse03}.

%
%
The temperature dependence of the spin decoherence time ($T_2$)
has been investigated using pulsed EPR at 240 GHz. The spin
decoherence time was measured by a Hahn echo sequence
($\pi/2-\tau-\pi-\tau-echo$) where the delay $\tau$ is
varied~\cite{schweiger}. The width of the pulses was adjusted to
maximize the echo signals and was typically between 200 ns and 300
ns. Because the corresponding excitation bandwidth of the applied
pulses ($\sim$0.15 mT width) was much smaller than the EPR
linewidth, a very small portion of Fe$_8$ spins was actually
manipulated in the $T_2$ measurement. Fig.~2(A) shows echo signals
with different delays and the echo area as a function of the delay
for the transition of $m_S$ = $-10$ $\leftrightarrow$ $-9$ at $T$
=1.27 $\pm$ 0.05 K. The spin decoherence time $T_2$ is estimated
from the decay rate of the echo area which is well fit by a single
exponential function, $exp(-2\tau/T_2)$ (Fig. 2(A)). The inset of
Fig.~2(A) shows the result of echo-detected field-sweep EPR at $T$
= 1.27 $\pm$ 0.05 K which shows the EPR transitions from $m_S$ =
$-10$ $\leftrightarrow$ $-9$. Like cw EPR, the echo-detected EPR
shows fine structures. Although the magnetic field was swept up to
12 T, no echo-detected EPR signals corresponding to other
transitions were observed. $T_2$ was measured between $T$ = 1.93
$\pm$ 0.05 K and 1.27 $\pm$ 0.05 K while a 4.566 T magnetic field
was applied along the easy axis. Above 1.93 K, $T_2$ became too
short to give spin echoes within the limited time resolution of
the pulsed spectrometer. Within this temperature range, the $T_2$
shows a strong temperature dependence and increases from $T_2$ =
93 $\pm$ 6 ns at 1.93 $\pm$ 0.05 K to $T_2$ = 714 $\pm$ 15 ns at
1.27 $\pm$ 0.05 K (see inset of Fig.~3). In addition, we measured
the temperature dependence of $T_2$ with the field orientation
along the hard plane which shows a similar temperature dependence
(not shown).

\begin{figure}
\includegraphics[width=85mm]{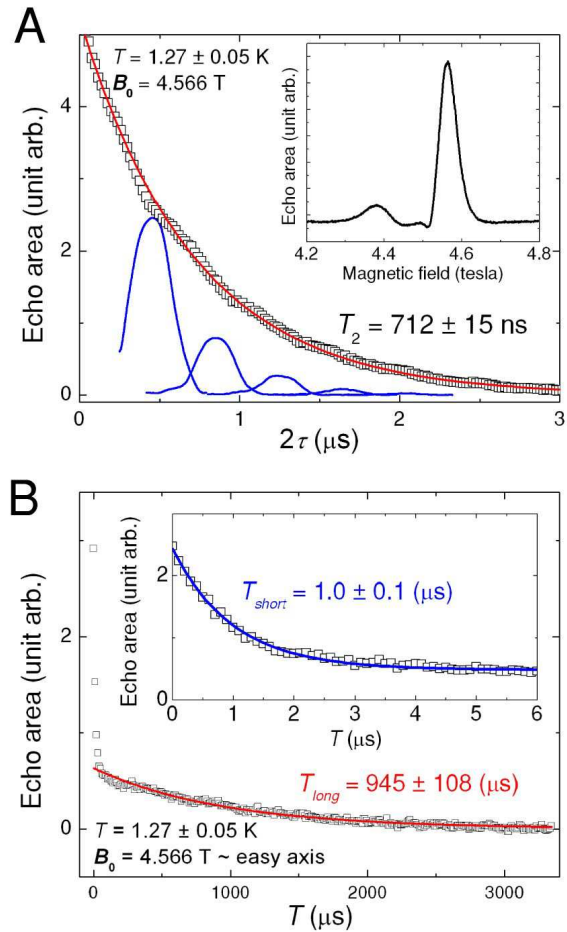}
\caption{\label{fig:invT1} (A) Echo signals and echo decay as a
function of 2{$\tau$}. The solid line is a fit using a single
exponential. Inset shows echo-detected EPR signals as a function
of magnetic field. (B) Echo decay measured by a stimulated echo
sequence. Inset shows a measurement of the echo decay in a short
timescale. Long decay curve was fit by a single exponential
$e^{-T/T_{long}}$ as shown in red solid lines while short decay
curve was fit by a double exponential $A e^{-T/T_{long}}+B
e^{-T/T_{short}}$ with $T_{long}$ = 945 $\mu$s.}
\end{figure}

The spin-lattice relaxation time $T_1$ was also investigated at
$T$ = 1.27 $\pm$ 0.05 K using a stimulated echo sequence
($\pi/2-\tau-\pi/2-T-\pi/2-\tau-echo$) where the delay $T$ is
varied~\cite{schweiger}. As shown in Fig.~2(B), we found two
relaxation rates in the echo decay curve with a short time of
$T_{short}$ = 1.0 $\pm$ 0.1 $\mu$s and a long time of $T_{long}$ =
948 $\pm$ 108 $\mu$s which both are longer than $T_2$. Because of
the small excitation bandwidth, a strong spectral diffusion is
expected. Therefore we currently speculate that the $T_{short}$ is
due to spectral diffusion and the $T_{long}$ is the spin-lattice
relaxation time $T_1$. However this value of $T_1$ is more than
two orders of magnitude longer than previous
findings~\cite{bahr07, bal08}. A more detailed investigation of
the $T_1$ processes will be presented elsewhere.

\begin{figure}
\includegraphics[width=95 mm]{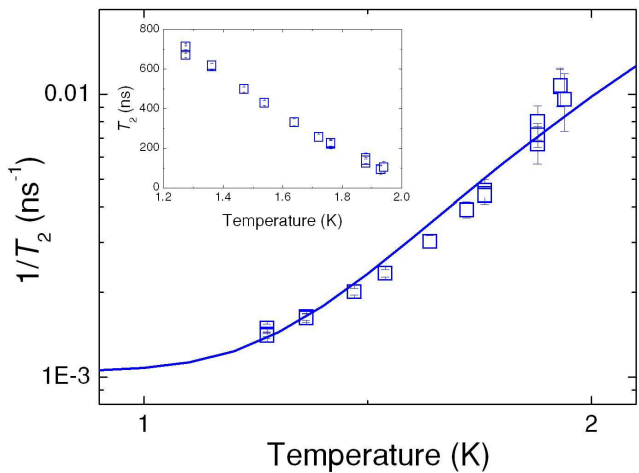}
\caption{\label{fig:invT2G} Temperature dependence of 1/$T_2$.
Data with error bars are shown by square dots and a simulation of
spin bath decoherence is shown by a solid line. The inset shows a
plot of $T_2$ vs. temperature.}
\end{figure}
Observation of a strong temperature dependence of $T_2$ suggests
that the main decoherence mechanism at higher temperatures is due
to dipolar coupling to fluctuating neighboring electron spins
which is often pictured as an electron spin
bath~\cite{takahashi08, Morello06}. Since the cw EPR spectra show
that Fe$_8$ SMMs dominate the population of electron spins in the
sample crystal, the source of the Fe$_8$ spin decoherence is
fluctuations of the Fe$_8$ spin bath itself which is formed by
intermolecularly dipolar-coupled Fe$_8$ SMMs. At $B_0$ = 4.6 T,
where the magnetic field is above all anti-level crossings and
single spin flips are suppressed, the Fe$_8$ spin bath fluctuation
is dominated by an energy-conserving spin flip-flop process. This
spin flip-flop rate is proportional to the number of pairs of
$m_S$ and $m_S \pm 1$ spins and therefore it strongly depends on
the spin bath polarization~\cite{takahashi08, kutter95}. At 1.27 K
and 4.6 T, 99 \% of the Fe$_8$ spin bath is polarized to the $m_S$
= $-10$ lowest lying state which reduces the spin flip-flop rate
significantly. Thus, this experiment tests if the main decoherence
mechanism of the Fe$_8$ SMM is the Fe$_8$ spin bath fluctuation.
To test the hypothesis of spin bath decoherence by other SMMs, we
extend the case of the two-level system~\cite{takahashi08,
kutter95} to a multi-level system. Here we obtain the relationship
between $T_2$ and the spin flip-flop rate by,
\begin{eqnarray}\label{eq:T2}
\frac{1}{T_2}=A\sum_{m_S=-10}^{9} W(m_S) P_{m_S} P_{m_S+1} +
\Gamma_{res}
\end{eqnarray}
\begin{eqnarray}\label{eq:T22}
P_{m_S}=\frac{e^{-\beta E(m_S)}}{Z}
\end{eqnarray} where $A$ is a temperature independent
parameter, $\beta$ = 1/($k_B T$), $Z$ is the partition function of
the Fe$_8$ SMM spin system and $\Gamma_{res}$ is a residual
relaxation rate which comes from other temperature independent
decoherence sources. The flip-flop transition probability with two
electron spins $W$($m_S$) is given by,
\begin{eqnarray}\label{eq:Wms}
W(m_S) = |<m_S+1, m_S|S_1^+ S_2^-|m_S, m_S+1>|^2 \nonumber\\+
|<m_S, m_S+1|S_1^- S_2^+|m_S+1, m_S>|^2.
\end{eqnarray}

Using the Eq.~(1)-(4), we plotted the temperature dependence of T2
as shown in Fig.~3. The proposed model is in good agreement with
the experimental data. Therefore the result strongly supports the
decoherence mechanism caused by the Fe$_8$ spin bath fluctuation.
The best agreement is obtained with $\Gamma_{res}$ = 1.0 $\times$
10$^{-3}$ ns$^{-1}$, which corresponds to $T_2$ = 1 $\mu$s. This
residual decoherence is likely related to hyperfine couplings to
$I=1/2$ for $^{57}$Fe (2.1 \% natural abundance) and $I=1/2$ for
proton nuclear moments in the molecules~\cite{wernsdorfer00}, and
represents the decoherence time one would expect in a highly
diluted single crystal of Fe$_8$ SMM. SMM crystals grown with
replaced by deuterium ($I=1$) and with isotopically pure $^{56}$Fe
($I=0$) may have even longer low-temperature decoherence times.
The given values of the $\Gamma_{res}$ and $A$ in Eq.~(2) also
lead to a few ns of the minimum $T_2$ around 10 K. Therefore this
indicates that the increase of $T_2$ by polarizing the spin bath
is more than two orders of magnitude.

%
%
In summary, the temperature dependence of the spin decoherence
time $T_2$ of Fe$_8$ SMMs was measured between 1.27 K and 1.93 K.
With increasing temperature, $T_2$ decreases by an order of
magnitude from 714 ns at 1.27 K. The identification of the main
and the second decoherence sources for the Fe$_8$ SMM and the
demonstration of the suppression of Fe$_8$ spin decoherence is
particularly important to engineer molecular magnets for future
quantum information processing applications. These measurements
also establish that high-frequency pulsed EPR at low temperatures
provides access to a frontier of spin decoherence in electron spin
systems.

%
%
This work was supported by research grants from NSF (DMR-0520481)
and the W. M. Keck Foundation (M.S.S., S.T., L.C.B and J.v.T.) and
NSF (DMR-0703925) (M.S.S.).

\end{document}